# A FRAMEWORK FOR CONTEXTUAL INFORMATION RETRIEVAL FROM THE WWW

Mr D.K. Limbu
Software Engineering Research Lab
Auckland University of Technology
Auckland, New Zealand
dilip.limbu@aut.ac.nz

Dr A.M. Connor
Software Engineering Research Lab
Auckland University of Technology
Auckland, New Zealand
andrew.connor@aut.ac.nz

Professor S.G. MacDonell
Software Engineering Research Lab
Auckland University of Technology
Auckland, New Zealand
stephen.macdonell@aut.ac.nz

**Abstract**

Search engines are the most commonly used type of tool for finding relevant information on the Internet. However, today's search engines are far from perfect. Typical search queries are short, often one or two words, and can be ambiguous therefore returning inappropriate results. Contextual information retrieval (CIR) is a critical technique for these search engines to facilitate queries and return relevant information. Despite its importance, little progress has been made in CIR due to the difficulty of capturing and representing contextual information about users. Numerous contextual information retrieval approaches exist today, but to the best of our knowledge none of them offer a similar service to the one proposed in this paper.

This paper proposes an alternative framework for contextual information retrieval from the WWW. The framework aims to improve query results (or make search results more relevant) by constructing a contextual profile based on a user's behaviour, their preferences, and a shared knowledge base, and using this information in the search engine framework to find and return relevant information.

## 1 INTRODUCTION

The Internet is, in its simplest terms, a huge, searchable database of information reached via a computer [1]. It makes available an enormous amount of information, the challenge then being one of finding *relevant* information [2]. Even the most experienced searchers find it difficult to identify and retrieve relevant information from the worldwide web (WWW) [3].

As useful as they are, today's search engines are far from perfect. Typical search queries are short and are often ambiguous, potentially returning inappropriate results. Including additional search terms can help to refine the search queries, but it is difficult for even experienced searchers to select the optimum query terms so that the desired subset of information is retrieved [4]. Moreover, these search engine results are based on simple keyword matches without any concern for the information needs of the user at a particular instance in time [5]. A critical goal of successful information retrieval on the web, then, is to identify which pages are of high quality and relevance to a user's query [6].

The need to better target a search on the information that will satisfy a user's information needs [4] is well recognised. In this regard today's search engines are lacking a personalisation mechanism that can 'understand' the query or reflect the information needs of a user at a particular instance in time and return customised results [5]. To provide the desired information to the user requires effective methods for identifying the user's task context, based on available information [7]. Contextual information retrieval has been and remains one of the major long-term challenges in information retrieval [8].

This paper discusses an alternative framework for contextual information retrieval from the WWW. The framework, under development at the Auckland University of Technology, has as its primary goals:

1. To develop/utilise technology to construct for each user a contextual profile, by combining user's behaviour, user's preferences and a shared knowledge base.

2. To develop/utilise technology to collect millions of users' contextual profiles from millions of machines.

3. To develop/utilise technology to define and construct shared knowledge that can be used to provide user feedback/suggestions and to refine search queries.

4. To integrate the outcomes of 1-3 above in a single framework. These features contribute to making this framework open, robust and scalable.

## 2 RELATED WORK

Information Retrieval (IR) is a wide, often loosely defined term and this paper only concern with Internet/Web information retrieval or Web searching. The Internet/Web information retrieval or Web searching, uses search engines that index a portion of the document as a full-text database or uses web directories, which classify selected Web documents by subject to search for relevant information [9].

Figure 1 illustrates the evolution of information retrieval, where retrieval of information from the web is a subset of general information retrieval methods. Contextual information retrieval for searching information on the web is not a new idea but has distinct challenges when compared to either general information retrieval or non-contextual information retrieval from the web.

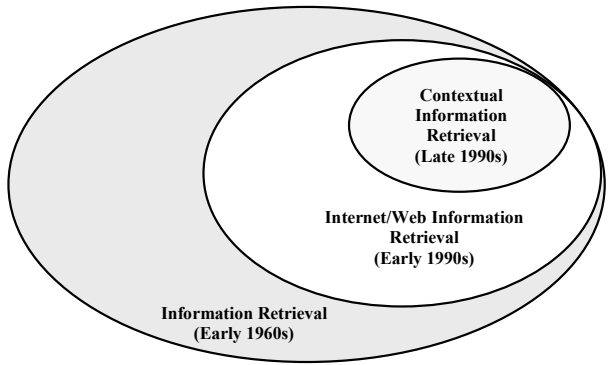

Figure 1: Evolution of Information Retrieval

There has been significant research in this area to date that has attempted to the overcome the major challenges of contextual information retrieval, and current research continues to improve the methods used. The key research in this area includes the development of PRISM [4], Letizia [10], the Wisconsin Adaptive Web Assistant (WAWA) [11] and Syskill & Webert [12].

Leake's PRISM [4] uses Watson [13] to monitor user behaviour in standard applications (such as word processors and Web browsers) and predict the type of information likely to be of interest to the user. Search queries are dispatched to special purpose search engines tailored towards the user's particular needs.

Lieberman's Letizia [10] monitors user's browsing behaviour, develops a user profile, and on that basis searches for potentially interesting pages for recommendation.

WAWA [11] constructs a Web agent by accepting user preferences in the form of instructions and adapting the agent's behaviour as it encounters new information. The system uses machine-learning methods to retrieve and/or extract textual information from the Web.

Pazzani et al.'s [12] Syskill & Webert asks the user to rank pages on a specific topic. Based on the content and rating of the pages, the system learns a user profile and predicts if pages encountered subsequently are likely to be of interest to the user.

Despite the achievements of these approaches, there remains no comprehensive model to describe the contextual information retrieval process [14] due to the difficulty of capturing and representing knowledge about users, context and tasks in a general Web search environment [8]. All of the above-mentioned approaches utilise either user behaviour or user preferences to construct a contextual profile, but not both. These approaches also do not use any form of shared intelligent knowledge base to provide relevance feedback or suggestions to the user and to formalise search queries. In addition, none of these approaches discuss how to use these captured contextual profiles in search engine server environments.

The remainder of this paper discusses an alternative framework for contextual information retrieval from the WWW that potentially addresses all the challenges mentioned above.

## 3  THE FRAMEWORK ARCHITECTURE

The proposed framework architecture is depicted in Figure 2. The architecture consists of two main models: Profile Collector and Context Manager, and it has three layers: Interface Layer, Knowledge Management Layer and Data Source/Search Engine Layer. Specialised autonomous agents reside at the various layers and perform well-defined functions. These agents support interactive monitoring and capturing of each user's behaviour and preferences, query specification and query processing, contextual profile gathering and categorisation, as well as relevance result filtering and presentation.

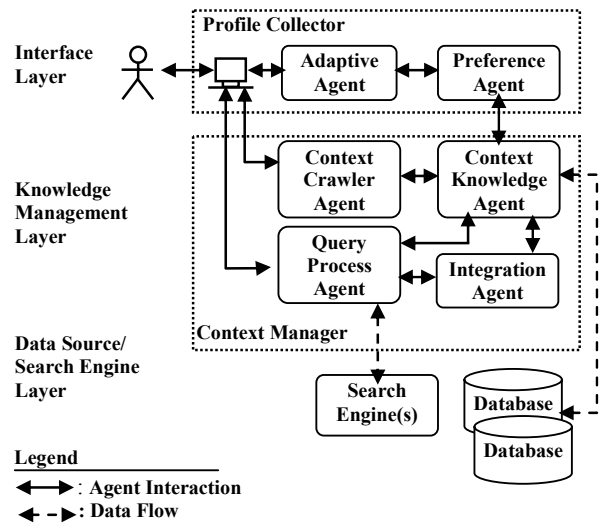

Figure 2: Proposed Framework Architecture

A simple usage scenario of the proposed framework (*as a system*) is given here. A user spent time at his/her desktop computer planning a holiday trip to New Zealand (*visiting travel web pages, booking airline tickets and making hotel reservation online; using Microsoft Word to store travel information, and sending emails to friends about trip.*). The system continuously monitors the user's desktop activities and captures user's contextual profile. When the user enters a query such as "Surfing", the system turns the "Surfing" keyword into shared concepts using the user's contextual profile and the existing knowledge based (uses various public ontology domains). The system then understands the meaning of "Surfing", i.e. "surfing waves" not *surfing the Internet*. In addition, the system also aware of the surfing location and surfing dates (*from the hotel address and booking date*). With these surfing information, the system generates contextual queries (*such as "Surfing in New Zealand", "Surf Tours", "Surf Lessons", "Surf Camps", "Surf Shops", "Northland surfing", "Auckland surfing", "East coast surfing" and so on*) and submits queries to a search engine. The system then filters results from the search engine using shared concepts and returns relevant information to user. The system also provides useful suggestions/feedbacks (*such as "Check weather", "Surfing guide", "See surfing pictures" and so on*) to user to get his/her specific interests.

The proposed architecture is general and modular so that new categorisations, ontologies and search engines can be easily incorporated. It also provides configurable features to store users' contextual profiles. Figure 3 shows how the proposed framework could be integrated with an existing search engine. The Profile Collector resides on the user's desktop and the Context Manager resides on the search engine server.

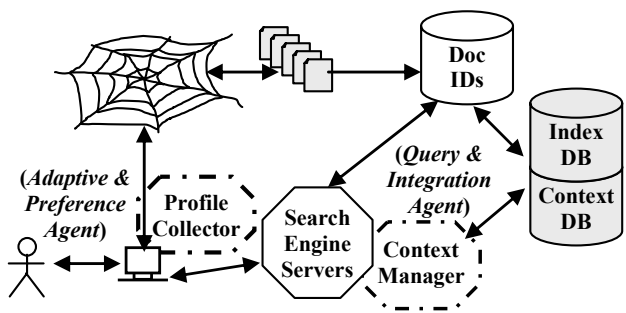

Figure 3: Proposed Framework Integrated with Search Engine

### 3.1 Profile Collector Model

The Profile Collector model has two agents, namely the Adaptive Agent and the Preference Agent. The agents reside on the user's desktop. They act as front-end brokers, gathering contextual information from the user. The Adaptive Agent maintains the user's contextual profile by monitoring and capturing the user's behaviour from their activities on the computer desktop. The Preference Agent interacts with the Adaptive Agent as well as the Context Knowledge Agent to provide the best possible suggestions/preferences to the user. The Preferences Agent also learns the user's preferences based on experience and feedback related to previous queries.

### 3.2 Context Manager Model

The Context Manager model has four agents: Context Crawler Agent, Context Knowledge Agent, Query Process Agent and Integration Agent.

The Context Crawler Agent is the most fragile agent since its role is to interact with millions of machines to gather users' contextual profiles. This is a challenging task as it raises performance and reliability issues. Perhaps more importantly, there are social implications arising from such a role (relating to privacy, spam, hacking, scams and so on). These issues warrant separate and extensive consideration. The Context Manager provides a configurable storage feature that enables a user to define where their contextual profile is stored: either on the local machine or in the search engine server. In the latter case, the user has to register or implicitly subscribe the context crawler service to store their profile on the server.

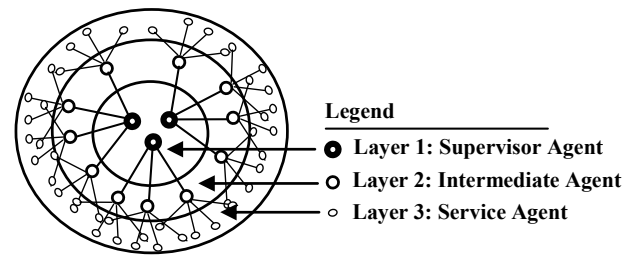

Figure 4: The Context Crawler Agent

The Context Crawler Agent in fact has three layers of agents as depicted in Figure 4, namely the Supervisor Agent, Intermediate Agent and Service Agent. The Supervisor Agent is a decision maker; the Intermediate Agent is a messenger between the Supervisor Agent and Service Agent; and the Service Agent gathers contextual profiles from millions of machines. Each layer agent has well-defined functions and these agents collaboratively gather users' contextual profiles. The Context Crawler Agent is an important avenue of investigation and it may well be a function of future search engines.

The Context Knowledge Agent is the most complex agent – it processes the millions of contextual profiles,

maintains the knowledge base, and queries various public ontology domains. The Context Knowledge Agent uses three dedicated agents to perform these tasks as depicted in Figure 5: the Context Processor Agent, Query Ontology Agent, and Knowledge Management Agent.

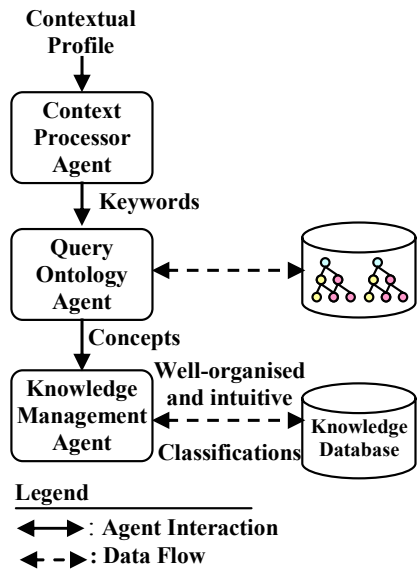

Figure 5: The Context Knowledge Agent

The Context Processor Agent processes and captures relevant keywords from each contextual profile. The Query Ontology Agent uses these keywords to create a shared understanding (or Keywords to Concepts) – between humans and for applications – information. The Knowledge Management Agent turns this information into well-organised and intuitive classifications and stores them for future use (such as to provide user feedback and recommendations and to refine search queries).

The Query Process Agent manages query formation and results management. The query formation task performs creation, manipulation, expansion, execution, persistence and reactivation of contextual queries. The results management task includes analysis, filtration, and relevance matching and formation of results from returned search results.

A query can be created with one or more terms supplied by the user. The Query Process Agent consults the Context Knowledge Agent to analyse the user's query syntax (phrase structure) and semantics and then decomposes the query into contextual sub-queries. This involves various processes: concept-based matching, disambiguation, lexical references, and query optimization. Alternatively, the user could start a query by picking one or more topics suggested by the Preference Agent. The Query Process Agent submits the contextual sub-queries or preference query to a search engine. Once results are returned, it processes them based on the user's contextual profile and sends them to the user.

The Integration Agent is a relatively simple agent that manages various public ontologies and search engines used by this framework. It communicates with the Query Process Agent and the Context Knowledge Agent to provide relevance information to support the whole contextual information retrieval process.

## 4 CONCLUSION

This paper has presented a research framework for contextual information retrieval from the WWW that will improve query results (or make search results more relevant). The proposed framework utilises various approaches/techniques to address some of the many acknowledged challenges that exist in the CIR domain.

The framework centres on the construction of user contextual profiles by combining user behaviour, user preferences and shared knowledge base information. The shared knowledge base can be used to provide user feedback/suggestions and to refine search queries. The framework requires the collection of millions of users' contextual profiles from millions of machines. Each of these components are then integrated in a single comprehensive CIR framework. These features contribute to making this framework open, robust and scalable

Our current work is focused on the investigation, specification, design, development and testing of the proposed framework. Once in place, we believe that the framework will be a significant contribution in Contextual Information Retrieval research as well as enhancing Information Retrieval in general.

## 5 REFERENCES

[1] McQuistan, S., "Techniques for current answers: Part 1: Information overload and the internet", Journal of Audiovisual Media in Medicine, vol. 23, no. 3, pp. 124, 2000

[2] Fan, W., Gordon, M.D. & Pathak, P., "Discovery of Context-Specific Ranking Functions for Effective Information Retrieval Using Genetic Programming", Knowledge and Data Engineering, IEEE Transactions, vol. 16, no. 4, pp. 523-527, 2004

[3] O'Hanlon, N., "Off the shelf & onto the Web: Web search engines evolve to meet challenges", Reference & User Services Quarterly, vol. 38, no. 3, pp. 247, 1999

[4] Leake, D.B. & Scherle, R., "Towards Context-Based Search Engine Selection", Proc. of the Intl. Conf. on Intelligent User Interfaces, 2001

[5] Challam, V.K.R., "Contextual Information Retrieval Using Ontology-Based User Profiles", Masters Thesis, Information and Telecommunication Technology Center, University of Kansas, 2004

[6] Sahami, M. *et al.*, "The Happy Searcher: Challenges in Web Information Retrieval", Proc. of the 8th Pacific Rim Intl. Conf. on Artificial Intelligence, 2004

[7] Bauer, T.L. & Leake, D.B., "Detecting Context-Differentiating Terms Using Competitive Learning", ACM SIGIR Forum, vol. 37, issue 2, pp. 4-17, 2003

[8] Allan, J., *et al.*, "Challenges in Information Retrieval and Language Modeling", ACM SIGIR Forum, vol. 37, issue 2, pp. 31-47, 2003

[9] Baeza-Yates, R. & Ribeiro-Neto, B. Modern Information Retrieval, ACM Press, 1999

[10] Lieberman, H. "Letizia: An Agent That Assists Web Browsing", Proc. of Intl Joint Conf. on Artificial Intelligence (IJCAI-95), 1995

[11] Eliassi-Rad, T. & Shavlik, J. "A System for Building Intelligent Agents that Learn to Retrieve and Extract Information", User Modeling and User - Adapted Interaction, vol. 13, no. 1-2, pp. 35, 2003

[12] Pazzani, M., Muramatsu, J. & Billsus, D. "Syskill & Webert: Identifying interesting web sites", Proc. of the National Conference on Artificial Intelligence (AAA1996). 1996

[13] Budzik, J. & Hammond, K.J. "User Interactions with Everyday Applications as Context for Just-in-time Information Access", Proc. of the Intl. Conf. on Intelligent User Interfaces (ICIUI2000), 2000

[14] Wen, J.-R., Lao, N. & Ma, W.-Y. "Probabilistic model for contextual retrieva", Proc. of the Annual ACM Conference on Research and Development in Information Retrieval, 2004